\documentclass[twocolumn,twocolappendix]{aastex631}
\usepackage{amsmath,amstext}
\usepackage[T1]{fontenc}
\usepackage[figure,figure*]{hypcap}
\usepackage [english]{babel}
\usepackage [autostyle, english = american]{csquotes}
\usepackage {booktabs}
\usepackage{subfigure}
\usepackage{amssymb}
\usepackage{multirow}
\usepackage{gensymb}
\usepackage{amsmath}
\usepackage{color}
\usepackage{hyperref}
\usepackage{marvosym}

\defcitealias{gao2019lithium}{Gao19LRS}
\defcitealias{gao2021lithium}{Gao21MRS}

\usepackage{float}

\usepackage{array}

\usepackage{natbib}
\usepackage{verbatim}
\usepackage{graphicx}
\usepackage[utf8]{inputenc}

\include{cmds}
\shortauthors{Cai et al.}

\begin{document}

\title{Li-rich Giants Identified from LAMOST DR8 Low-Resolution Survey}

\correspondingauthor{XiaoMing Kong}
\email{xmkong@sdu.edu.cn}

\author{BeiChen Cai}
\affiliation{School of Mechanical, Electrical \& Information Engineering, Shandong University, Weihai, 264209, Shandong, People’s Republic of China}

\author{XiaoMing Kong}
\affiliation{School of Mechanical, Electrical \& Information Engineering, Shandong University, Weihai, 264209, Shandong, People’s Republic of China}

\author{JianRong Shi}
\affiliation{Key Laboratory of Optical Astronomy, National Astronomical Observatories, Chinese Academy of Sciences, Beijing 100101, People’s Republic of China}

\author{Qi Gao}
\affiliation{Key Laboratory of Optical Astronomy, National Astronomical Observatories, Chinese Academy of Sciences, Beijing 100101, People’s Republic of China}

\author{Yude Bu}
\affiliation{School of Mathematics and Statistics, Shandong University, Weihai, 264209, Shandong, People’s Republic of China}

\author{Zhenping Yi}
\affiliation{School of Mechanical, Electrical \& Information Engineering, Shandong University, Weihai, 264209, Shandong, People’s Republic of China}

\begin{abstract}
A small fraction of giants possess photospheric lithium(Li) abundance higher than the value predicted by the standard stellar evolution models, and the detailed mechanisms of Li enhancement are complicated and lack a definite conclusion. In order to better understand the Li enhancement behaviors, a large and homogeneous Li-rich giants sample is needed. In this study, we designed a modified convolutional neural network model called Coord-DenseNet to determine the A(Li) of Large Sky Area Multi-Object Fiber Spectroscopic Telescope (LAMOST) low-resolution survey (LRS) giant spectra. The precision is good on the test set: $\rm MAE=0.15$\,dex, and $\sigma=0.21$\,dex. We used this model to predict the Li abundance of more than 900,000 LAMOST DR8 LRS giant spectra and identified 7,768 Li-rich giants with Li abundances ranging from 2.0 to 5.4\,dex, accounting for about 1.02\,\% of all giants. We compared the Li abundance estimated by our work with those derived from high-resolution spectra. We found that the consistency was good if the overall deviation of 0.27\,dex between them was not considered. The analysis shows that the difference is mainly due to the high A(Li) from the medium-resolution spectra in the training set. This sample of Li-rich giants dramatically expands the existing sample size of Li-rich giants and provides us with more samples to further study the formation and evolution of Li-rich giants.

\end{abstract}


\section{INTRODUCTION}
The element lithium is one of the three light elements produced by primordial nucleosynthesis after the Big Bang and exists mainly in the form of the isotope $ {}^{7}\mathrm {Li}$\,(Hereafter Li or lithium refers to $ {}^{7}\mathrm {Li}$). Its abundance is the most reliable way to study the physical conditions of the universe in the initial period after the Big Bang\citep{steigman2007primordial,khatri2011time,fields2020big,romano2021gaia}. According to the standard stellar evolution theory, the atmospheric Li abundance of a star with around Solar mass will be largely preserved throughout the star's main-sequence because of the relatively low temperature in the atmosphere\citep[]{sweigart1979meridional}. Then, as the star evolves from the main sequence toward the red giant branch (RGB) stage, the first dredging process occurs\citep[]{iben1965stellar}, the convective envelope deepens and the surface lithium is transported to the hot stellar interior, where the temperature (about 2.6 million Kelvins, \citealt{gamow1933internal,salpeter1955nuclear}) is high enough to cause the lithium element to be consumed severely. Many studies have suggested that when a star with approximately solar mass finishes its first dredge-up process, its surface Li abundance should be A(Li)\footnote[4]{
  $A_{\mathrm{Li}}=\log \left(\mathrm{N}_{\mathrm{Li}} / \mathrm{N}_{\mathrm{H}}\right)+12$} 
  $\leq1.5$ dex\,\citep[]{brown1989search,lind2009signatures,liu2014lithium}.

However, \citet{wallerstein1982ak} discovered a K-giant with A(Li)\,\textgreater\,3.2 dex, which is not well explained by the standard stellar evolution model. Since then, more and more Li-rich giants have been identified\citep[e.g., ][]{hanni1984lithium,brown1989search,gratton1989hd,lind2009signatures,kirby2012discovery,martell2013lithium,adamow2014penn,casey2016gaia,smiljanic2018gaia,carbon2018search,zhou2019high,gao2019lithium,gao2021lithium,2020MNRAS.498...77H}. Based on the obtained observation data, the proposed scenarios to explain the mechanism of lithium acquisition or production during stellar evolution include: the extra mixing in the stellar interior \citep[e.g.,][]{sackmann1992creation,charbonnel2000nature,zhou2019high}; engulfment/accretion of planets containing Li \citep[e.g.,][]{alexander1967possible,siess1999accretion,carlberg2012observable,aguilera2016lithium}; the merger between a red giant star and a companion helium white dwarf \citep[]{holanda2020tyc,zhang2020population}; tidal interactions between binary stars which drive Li production in low-mass red giants \citep{casey2019tidal}. Several recent studies \citep{reddy2019study,singh2019survey,yan2021most} have found that red clump\,(RC) stars have more Li-rich giants than RGB stars. There is also a fraction of Li-rich giants that exhibit infrared\,(IR) excess \citep[]{de1996lithium,rebull2015infrared}, but they are not systematically different from normal giants with the same stellar parameters and evolutionary stage \citep{martell2013lithium,casey2016gaia,smiljanic2018gaia,martell2021galah}.

Over the decades, many different methods have been applied to find Li-rich giants to better understand the mechanisms of lithium enrichment during stellar evolution. For example, \citet{martell2013lithium} identified 23 Li-rich giants by the spectral index method. \citet{kumar2018identifying} identified 15 Li-rich K giants by the line core intensity ratio of the Li line at 6707{\AA} to the Ca line at 6717{\AA}. \citet[hereafter \citetalias{gao2019lithium}]{gao2019lithium} identified 10,535 Li-rich giants using the template matching method from the  Large Sky Area Multi-Object Fiber Spectroscopic Telescope(LAMOST, \citealp{cui2012large,zhao2012lamost}) survey. Although the sample size of Li-rich giants has been greatly extended by these works, the fraction of Li-rich giants is very small compared to the number of normal stars. Most studies believe that Li-rich giants account for about $0.5\,\%-1.5\,\%$ of all giants\citep[e.g.,][]{brown1989search,gonzalez2009li,kumar2011origin,ruchti2011metal,lebzelter2012lithium,liu2014lithium,casey2016gaia,gao2019lithium,smiljanic2018gaia}. Unlike the former, \citet{adamow2014penn} believe that the proportion is about 2$\,\%$. In addition, some studies suggest that the ratio is about $0.2\,\%-0.3\,\%$\citep{martell2013lithium,reddy2019study}. A large homologous sample set is necessary to better understand the evolutionary mechanisms of 
Li-rich giants and the tens of millions of spectra published by the LAMOST survey provide an excellent opportunity to systematically identify Li-rich giants.

On March 31, 2021, the LAMOST DR8 dataset, which contains the pilot survey and the first eight years of the official survey, was officially released to domestic astronomers and international collaborators. The released DR8 dataset consists of two parts: regular low-resolution spectra data and medium-resolution spectra data, including 5207 low-resolution observational sky areas and 1089 medium-resolution observational sky areas. The total number of released spectra reached 17.23 million, including 11.21 million low-resolution spectra, 1.47 million medium-resolution non-time-domain spectra, and 4.55 million medium-resolution time-domain spectra. Among them, 13.28 million spectra with DR8 signal-to-noise ratio greater than 10 were released. In addition, DR8 also released the stellar spectral parameters of 7.75 million stars, which is currently the largest catalog of stellar spectral parameters in the world.

With the advent of the era of astronomical big data, the application of machine learning algorithms in this field has also been rapidly developed. In the field of astronomical research,
\citet{li2018carbon} used the Bagging TopPush algorithm to identify carbon stars in LAMOST DR4. \citet{bu2019searching} presented a framework combining convolutional neural networks\,(CNN, \citealt{lecun1998gradient}) and support vector machine\,(SVM) for classifying hot subdwarf stars from LAMOST DR4. \citet{leung2019deep} designed a neural network and used it for the analysis of APOGEE high-resolution spectra. \citet{yi2019efficient} used the XGBoost algorithm to search for M giants in LAMOST DR5. \citet{wang2020spcanet} design a new structure for the network SPCANet based on CNN to estimate the fundamental stellar atmospheric parameters\,($\mathrm{\emph{\rm T}_{eff}}$ and $\rm log\,{g}$)  and  13 chemical abundances of 1,472,211 spectra from LAMOST-\uppercase\expandafter{\romannumeral2} medium-resolution spectroscopic survey (MRS). \citet{sun2021exploring} trained a SLAM model based on support vector regression and used it to predict the fundamental stellar parameters of more than 40,000 late-B and A-type main sequence stars from LAMOST DR7. In this paper, we construct a data-driven model for LAMOST low-resolution spectra based on CNN called Coord-DenseNet, and then use it to systematically search for potential Li-rich giants from the low-resolution spectra of LAMOST DR8. We provide a catalog of recognized Li-rich giants with the LAMOST ID, position, surface gravity,  metallicity, effective temperature, and Li abundance of the stars.

The paper is organized as follows: in Section \ref{sec:DATA}, we briefly introduce the data used in this paper. Section \ref{sec:Method} describes the structure of the Coord-DenseNet. Section \ref{sec:Experiment} gives the evaluation method of the model and the performance on the test set.
  Section \ref{sec:Discussion} compares the A(Li) predicted by the model with those given in the literature and gives a discussion of the properties of the predicted Li-rich giants. Section \ref{sec:Summary} gives a brief summary of the paper.

\section{DATA}\label{sec:DATA}

\subsection{Spectra} 

LAMOST is a reflecting Schmidt telescope with a combination of a large aperture and a large field of view. Its focal plane is circular with a diameter of 1.75 m ($\sim$5°) and 4000 fibers are evenly distributed on it, which allows it to obtain 4000 spectra in a single exposure.
LAMOST has two modes, with a resolution of R$\sim$1800 in the low-resolution mode and R$\sim$7500 in the medium-resolution mode. The LAMOST  DR8 low-resolution survey (LRS) contains a total of 11,214,076 wavelength-calibrated and relative flux-calibrated spectral data from LAMOST observations from October 24, 2011, to May 27, 2020, including 10,388,423 stellar spectra covering the wavelength range from  3690{\AA} to 9100{\AA}.

\begin{figure}[hbt]
\centering
\includegraphics[width=0.48\textwidth]{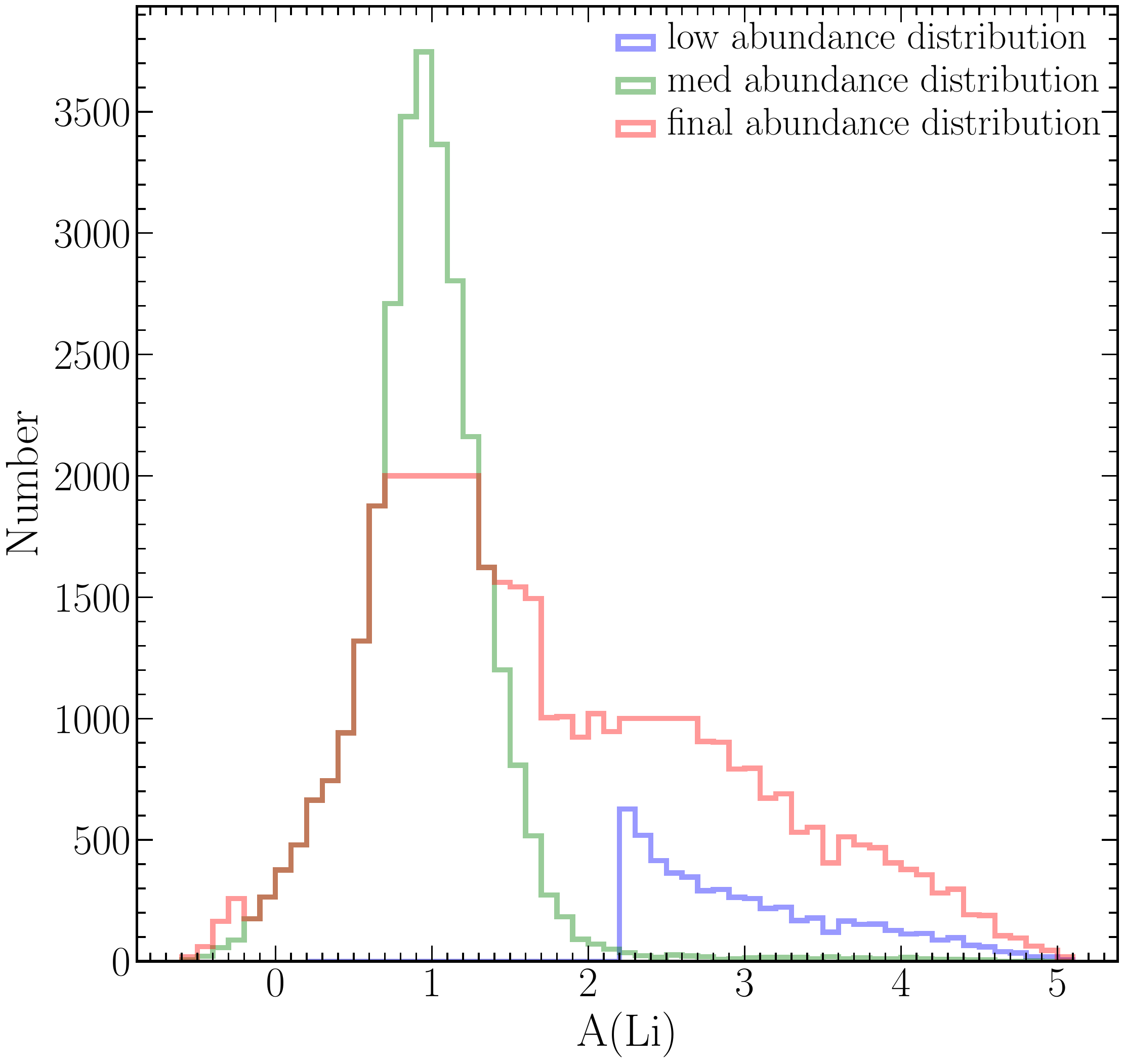}
\caption{The distribution of Li abundance in the sample dataset. The bin size is 0.1dex. The abundances of the 30,445 spectra in the green solid line come from \citetalias{gao2021lithium}, the abundances of the 5545 spectra in the solid blue line are corrected by Eq.\ref{eq:1} and data augmented, the abundance before correction is provided by \citetalias{gao2019lithium} and the red solid line is the abundance distribution of 44,959 spectra adopted as final training samples after data augmentation and undersampling.}\label{fig:fig1}
\end{figure}

\begin{figure*}[!t]
\centering
\includegraphics[width=0.9\textwidth]{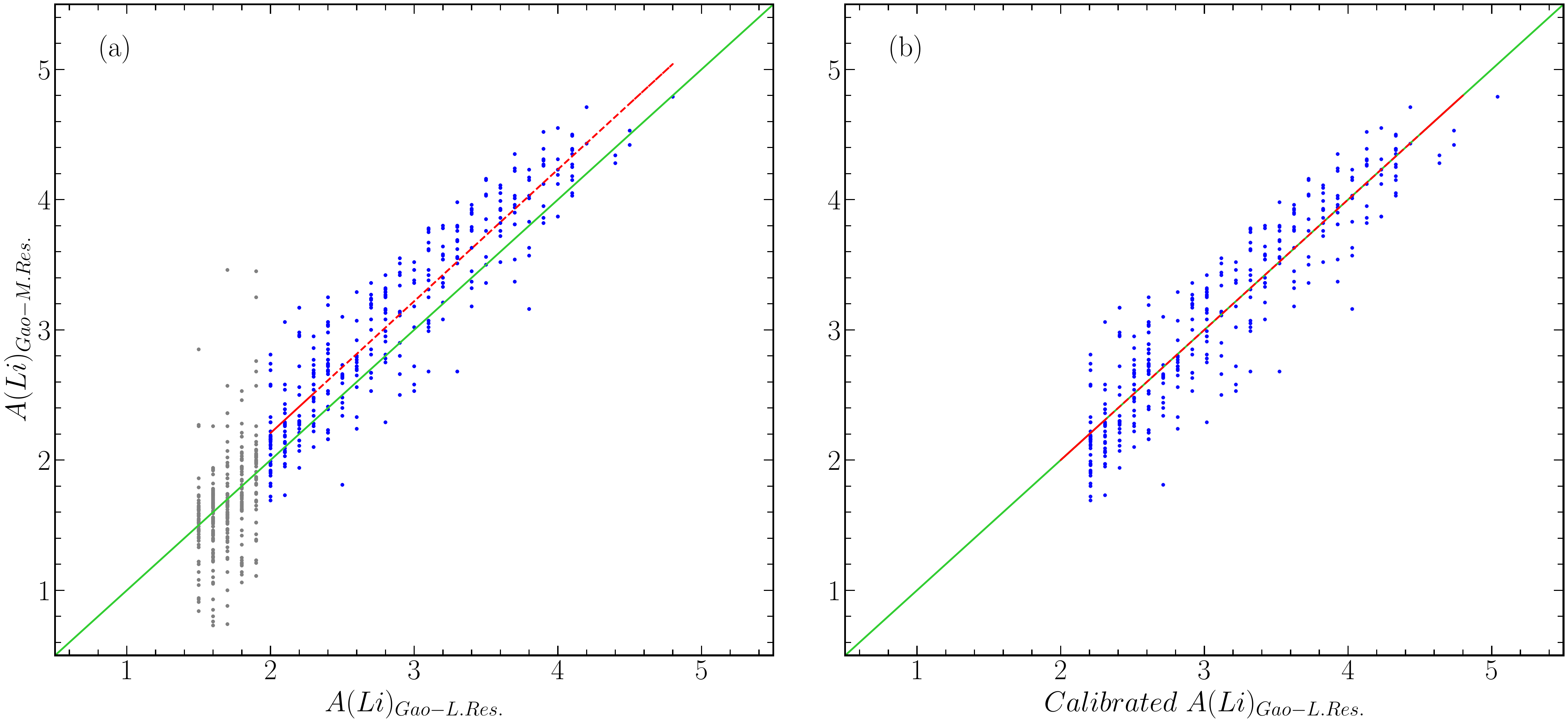}
\caption{  \emph{Panel\,(a):} Comparison of the $\rm A(Li)_{M.Res.}$ from the medium-resolution spectra of \citetalias{gao2021lithium} and the $\rm A(Li)_{L.Res.}$ from the low-resolution spectra of \citetalias{gao2019lithium} in the dataset, the gray points are the fractions with $\rm A(Li)_{L.Res.} \textless 2$ dex, the blue points are the fractions with $\rm A(Li)_{L.Res.} \geq 2$ dex, the green solid line is the diagonal line, and the red dashed line is the best linear fitting of the blue points. \emph{Panel\,(b):}
The blue dots are the $\rm A(Li)_{M.Res.}$ compared to the calibrated $\rm A(Li)_{L.Res.}$, the red dashed line is the best linear fitting of the blue dots.
\label{fig:2}}
\end{figure*}

\subsection{Data sets} 

We plan to search for Li-rich giant stars in LAMOST DR8 LRS. To achieve this goal, we used label samples to train a lithium abundance prediction model that predicted the Li abundance of DR8 LRS giants spectra and then screened Li-rich samples for further analysis. The spectra of a dataset in this paper are from LAMOST DR8 LRS, while the Li abundance as a sample label consists of two components:
\begin{itemize}
\item [1.] 
Li abundance derived from LAMOST medium-resolution spectra by \citet[hereafter \citetalias{gao2021lithium}]{gao2021lithium}. \citetalias{gao2021lithium} applied the template matching method to obtain the Li abundance of 165,479 stars corresponding to 294,857 spectra based on LAMOST medium-resolution spectra, and we used the $\rm log\,{g}$ $\textless$ 3.5 and $\mathrm{\emph{T}_{eff}}$ $\textless$ 5600K criteria to filter the Li abundances of 42,346 giants corresponding to 71,208 spectra. Then we cross-match this catalog with LAMOST LRS DR8 by coordinating and getting 30,373 giants corresponding to 30,445 low-resolution spectra. As shown in the green line graph in Figure\,\ref{fig:fig1}.
\item [2.] 
As can be seen in Figure\,\ref{fig:fig1}, the sample labels from the medium resolution are mainly distributed in the range of 0 $\textless$ A(Li) $\textless$ 2.0\,dex, and the number of A(Li) $\textgreater$ 1.5\,dex samples is too small, especially the number of samples with A(Li) $\textgreater$ 2\,dex. \citetalias{gao2019lithium} used the template-matching method to search for 10,535 giants with an abundance ranging from 1.5\,dex to 4.9\,dex from LAMOST DR7 LRS. To increase the prediction accuracy of the model in this part, we have used the star catalog released by \citetalias{gao2019lithium} as a supplement. In order to compare the consistency of the two sets of label samples, we cross-matched the low-resolution and medium-resolution catalogs published by \citepalias{gao2019lithium,gao2021lithium} and obtained a total of 697 spectra, as shown in Figure\,\ref{fig:2}. As can be seen from Figure\,\ref{fig:2}(a), For the 1.5\,dex $\textless$ $\rm A(Li)_{Gao-L.Res.}$ $\textless$ 2.0\,dex part, the dispersion is a bit larger and a considerable part of the samples with A(Li) $\textless$ 1.5\,dex in the medium resolution spectra are considered to be greater than 1.5 dex in low resolution. We analyzed that in the region of 1.5\,dex $\textless$ A(Li) $\textless$ 2.0\,dex, the absorption line of Li \uppercase\expandafter{\romannumeral1} is generally weak and the depth change is not obvious, which is easy to cause large errors for low-resolution spectra with low S/N. On the other hand, for the A(Li) $\geq$ 2\,dex part, the dispersion is smaller but the data obtained from the medium-resolution spectra is obviously higher than the overall data obtained from the low-resolution spectra. After comprehensive consideration, we used the fraction of the low-resolution spectra with A(Li) $\geq$ 2\,dex, and after removing a few outliers with abundance dispersion greater than 1\,dex, a linear equation was obtained:
\begin{equation}\label{eq:1}
A(\rm Li)_{M.Res.}=1.01 \times A(\rm L i)_{L.Res.}+0.18
\end{equation}
We use the $\rm A(Li)_{M.Res.}$ as a benchmark to correct for the $\rm A(Li)_{L.Res.}$ according to Eq.\ref{eq:1}, see Figure\,\ref{fig:2}(b). For the low-resolution spectra of 697 lithium giants from \citetalias{gao2019lithium}, we adopted the part with A(Li) $\geq$ 2\,dex and corrected its A(Li) by Eq.\ref{eq:1}. Finally, we got 5545 label samples, as shown in the blue solid line in Figure\,\ref{fig:fig1}.
\end{itemize}

In order to obtain a high-quality dataset, the data were further filtered by the following conditions:
\begin{itemize}
\item [1)]
Spectra with S/N $\textless$ 30 pixels in the r-band were excluded.
\item [2)]
The spectra with $\rm redshift = - 9999$ provided by LAMOST was excluded, which means the red-shift of these spectra was not measured accurately.
\item [3)]
For spectra with both low-resolution spectra and medium-resolution spectra Li abundance values, the A(Li) from medium-resolution spectra was adopted.
\end{itemize}

As shown in Figure\,\ref{fig:fig1}, the number of samples in different A(Li) regions is extremely uneven, and in the region, A(Li) $\textgreater$ 1.5\,dex, the number of samples decreases extremely rapidly with the increase of A(Li). In order to ensure the generalization ability of the model, we reduce the sample size gap in different A(Li) regions by data augmentation \citep{oh2020time} and undersampling. First, the number of data within the bins of A(Li) from 0.7\,dex to 1.3\,dex at 0.1\,dex intervals is undersampled to 2000, and then the number of samples is increased to about three times the original number by first-order interpolation for A(Li) $\textgreater$ 1.6 dex and A(Li) $\textless  -0.2$ dex parts of the data. Taking triple upsampling as an example, the basic principle of the  interpolation method is:
\begin{itemize}
\item [1)]
The original data is a one-dimensional array of fixed length and a spectral fitting curve is obtained by first-order interpolation of the two adjacent elements in the array.
\item [2)]
Resample the flow values on the fitted curve at a density three times the length of the original array to obtain a new array three times the length of the original array.
\item [3)]
Start from the first, second, or third element of the new array, take one of every three elements, and get 3 arrays with the same
length as the original array, which is the new spectral data obtained after triple upsampling.
\end{itemize}

\begin{figure*}[!t]
\centering
\includegraphics[width=0.9\textwidth]{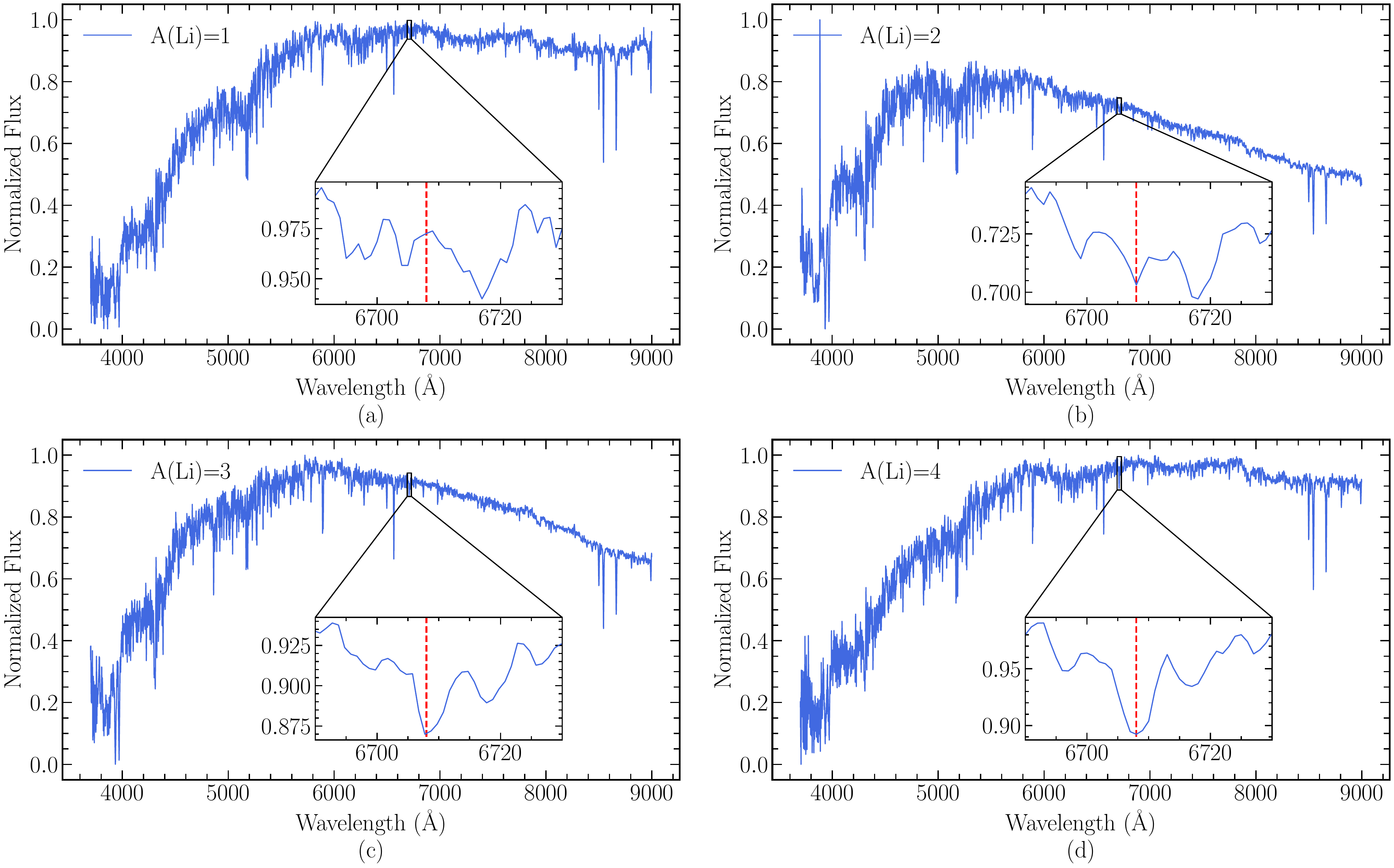}
\caption{Example spectra of LAMOST LRS with different A(Li). the red dashed line in the figure is the absorption line of Li at 6708 {\AA}. The four subplots (a), (b), (c), and (d) are the graphs of the absorption lines for A(Li) of 1\,dex, 2\,dex, 3\,dex, and 4\,dex, respectively. It can be seen that as the A(Li) increases, the absorption lines become deeper.
\label{fig:3}}
\end{figure*}

\begin{figure}[!t]
\centering
\includegraphics[width=0.49\textwidth]{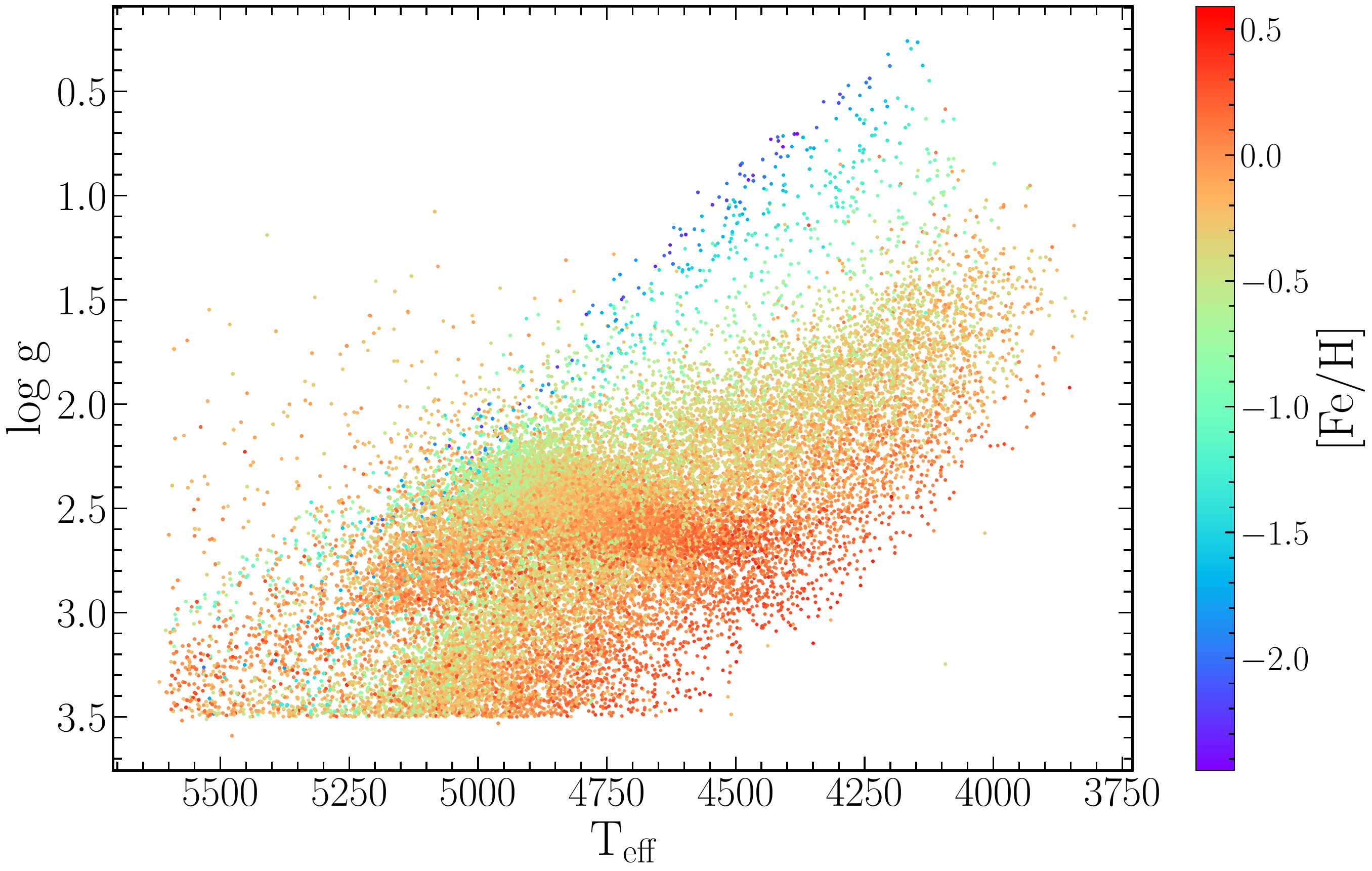}
\caption{Distribution of $\rm T_{eff}$ and $\rm log\,{g}$ for the data set colored by [Fe/H].}\label{fig:4}
\end{figure}

The final sample size of the dataset
after data augmentation and undersampling is 44,959. An illustration of the spectra used in the experiments is shown in Figure\,\ref{fig:3}. Figure\,\ref{fig:4} shows the stellar parameter space of our sample in the panel of  $\rm T_{eff} - log\,{g}$.

\subsection{Data pre-processing} 

Before starting to train the model, we preprocessed the data is as follows:


To eliminate the effect of different scales of flux values of different spectra on model training, The flux of all the spectra were  normalized to the range [0,1] by the following equation: 
\begin{equation}
\operatorname{Flux}_{\text {Norm }}=\frac{\operatorname{Flux} -\operatorname{Min}(\operatorname{Flux} )}{\operatorname{Max}(\operatorname{Flux} )-\operatorname{Min}(\text { Flux })}
\end{equation}

where $\operatorname{Flux}_{\text {Norm }}$ is the normalized flux, $\rm Flux$ is the flux of an observed spectrum, $\operatorname{Min}(\rm Flux)$ is the minimum  flux for the spectrum, and $\operatorname{Max}(\rm Flux)$ is the maximum flux for that spectrum. After normalization, the fluxes of all the spectra are in the range of $[0-1]$.

The processed data are randomly divided into a training set, validation set, and test set in the ratio of 7:2:1. The training set is used to train the DenseNet model, and the validation set is used several times during the model training process to continuously adjust the hyperparameters (e.g., learning rate, number of network layers, size of the convolutional kernel, etc.) and to determine whether the model is overfitted. The test set evaluates and verifies the performance of the model after the model training is completed. When the loss function of the model converges and stabilizes, the test set is used for the final test, and the error of this time is used as an approximation of the generalization error.

\begin{figure*}[htb]
\centering
\includegraphics[width=0.9\textwidth]{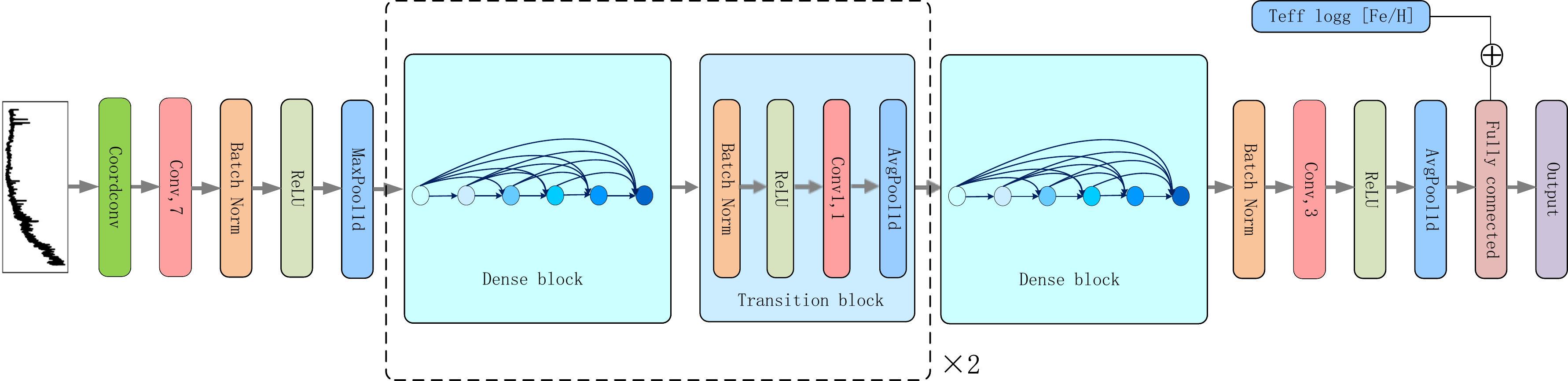}
\caption{The network structure of the DenseNet model. The arrows indicate the direction of data flow in the network.
\label{fig:5}}
\end{figure*}

\begin{figure*}[htb]
\centering
\includegraphics[width=0.9\textwidth]{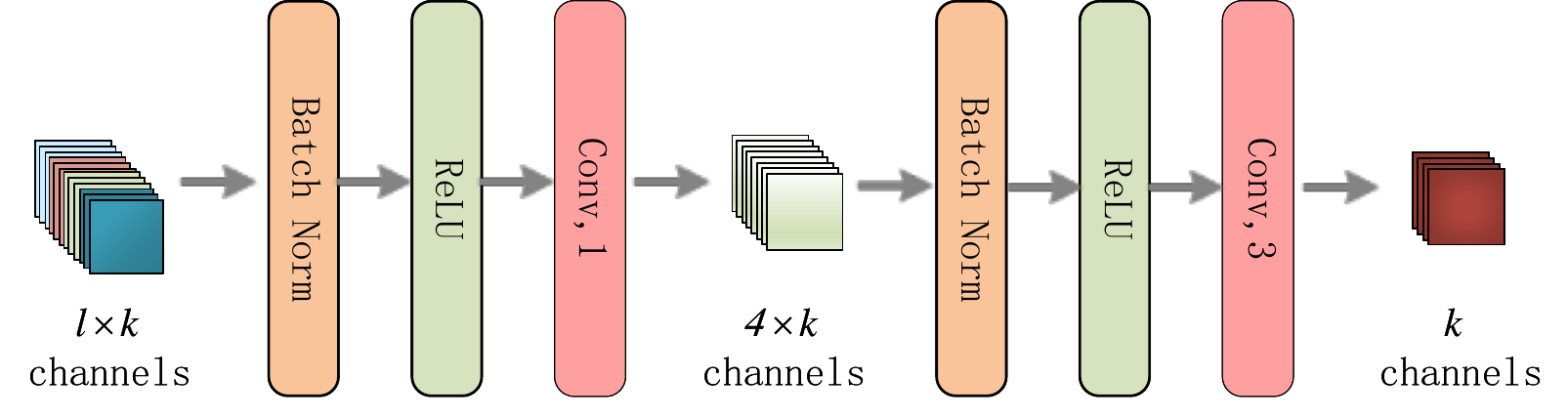}
\caption{The structure of the dense layer. Since the dense layer accepts the output of the previous layer as input, the input of the later layer will be very large. The role of the convolutional layer with a kernel size of 1 in the dense layer is to change the $l \times k$ channel into a $4 \times k$ channel, so as to achieve the purpose of reducing the number of features and improving the computational efficiency.
\label{fig:6}}
\end{figure*}

\section{Method} \label{sec:Method}
DenseNet is a CNN-based deep learning model proposed by \citet{huang2017densely} for image classification in the field of computer vision. As a deep CNN, DenseNet is suitable for feature processing of high-dimensional data. Compared with the traditional CNN, the advantages of DenseNet are mainly as follows:
\begin{itemize}
\item [1)] 

Due to the dense connection between the layers, DenseNet enhances the back-propagation of gradients, which makes the network easier to train. The layers are implicitly deep and supervised by short connection.
\item [2)]
DenseNet achieves feature reuse through the concatenation of feature maps, making it less parametric and more computationally efficient.
\item [3)]
It alleviates the gradient disappearance and gradient explosion problems that arise when the layers of a neural network are deep.
\end{itemize}
The original DenseNet is used for 2D image classification tasks and cannot be directly used for 1D spectra. We improved the original DenseNet to make it applicable to 1D spectra\,(see Figure\,\ref{fig:5}).

Convolution has translation invariance so that the parameters of the convolution kernel can be shared uniformly at different locations of the image. However, when the convolution performs local operations, it doesn't know the spatial location where the current convolution kernel is located. Unlike the picture classification task, similarly shaped absorption lines at different wavelengths have completely different meanings for the spectrum. For this problem, we introduce Coordconv\citep{liu2018intriguing} as a solution by adding a Coordconv layer at the beginning of the model. Coordconv generates a hard-coded position feature map and concatenates it with the input feature map in the channel dimension. At this point, CoordConv has a certain translation dependency. Then the traditional convolution is performed so that the convolution process can perceive the spatial information of the feature map.

The Coord-DenseNet model we use as a kind of CNN mainly consists of a 1D convolutional layer\,(Conv), 1D pooling layer, 1D batch normalization layer\,(BN), and fully connected layer\,(FC). The convolutional layer is the core of the neural network, and the convolutional operation can break the limitation of traditional filters to extract the desired features according to the objective function. Each convolutional layer is followed by a nonlinear activation unit\,(ReLU), which does a nonlinear mapping of the output of the convolutional layer. The pooling layer can be divided into an average pooling layer\,(AvgPool) and a maximum pooling layer\,(MaxPool) according to the calculation method, and the pooling layer can reduce the dimension of the feature information extracted from the convolutional layer. On the one hand, it can make the feature map smaller, simplify the computational complexity of the network and avoid overfitting to a certain extent; on the other hand, it can compress the features and retain the significant features. As the network deepens or during the training process, the distribution of data will be gradually shifted, which will lead to the disappearance of a gradient in the lower layer of the neural network when back-propagation, the role of the BN layer is to force the skewed distribution into a standard normal distribution, which can effectively avoid the problem of gradient disappearance and accelerate the convergence speed of the neural network. The fully connected layer integrates the highly abstracted features learned from previous convolutions for the final classification or regression.

The main framework of Coord-DenseNet consists of three dense blocks and two transition blocks. Except for the last dense block, each dense block is followed by a transition block. Each dense block includes six densely connected dense layers, and the structure of the dense layers is shown in Figure\,\ref{fig:6}. In each dense block, each dense layer accepts the output of all the previous dense layers as its own input. This connection strengthens the connection between the features extracted from different convolutional layers and also alleviates the gradient disappearance problem encountered in the deep neural network during training. The use of more dense layers in the experiment cannot further improve the accuracy of the model while making the computational overhead increase dramatically, while fewer dense layers make the model unable to achieve the existing accuracy.

The structure of the transition block is a BN layer, a ReLU function, a 1D convolutional layer with a kernel size of 1, and a 1D AvgPooling layer. The main role of the transition block is to connect two adjacent dense blocks and reduce the feature map size. In addition, the transition block can also serve to compress the model. Suppose the number of channels of the feature map output from the Dense block before the transition is $m$, then it will become $\left \lfloor \theta m \right \rfloor$ feature maps after the transition block, where $\theta  \in \left( {0,1} \right]$ is the compression ratio.

Li abundance estimates are usually based on  Li \uppercase\expandafter{\romannumeral1} resonance line at 6708 Å. In order to reduce the number of parameters and the training time of the model, we only took the segment of the spectrum at 6660\,Å$\,-\,$6740\,Å as the data input to the model, and since the estimation of Li abundance is closely related to the atmospheric parameters of the spectrum, we concatenated $\rm{T_{eff}}$, $\rm log\,{g}$, [Fe/H] provided by LAMOST with the fully connected layer of the model to predict the A(Li) of the spectrum.

\setlength{\tabcolsep}{1mm}{
\begin{table}[htbp]
	\centering
    \caption{Number of Coord-DenseNet Predicted Results on Target Value A(Li) $\textless$ 1.5\,dex and A(Li) $\textgreater$ 1.5\,dex}
	\label{table1}  
	\begin{tabular}{cccc ccc}
		\hline\hline\noalign{\smallskip}
                                      &            & \multicolumn{2}{c}{A(Li) }  \\ 
\cline{3-4}
                                      &            & A(Li) $\textless$ 1.5 & A(Li) $\textgreater$ 1.5          \\ 
\hline
\multirow{2}{*}{A(Li) (predict)} & A(Li) $\textless$ 1.5 & 2157       & 96                  \\
                                      & A(Li) $\textgreater$ 1.5 & 56         & 2170                \\
\hline
\end{tabular}
\end{table}
}

\begin{figure}[htb]
\centering
\includegraphics[width=0.46\textwidth]{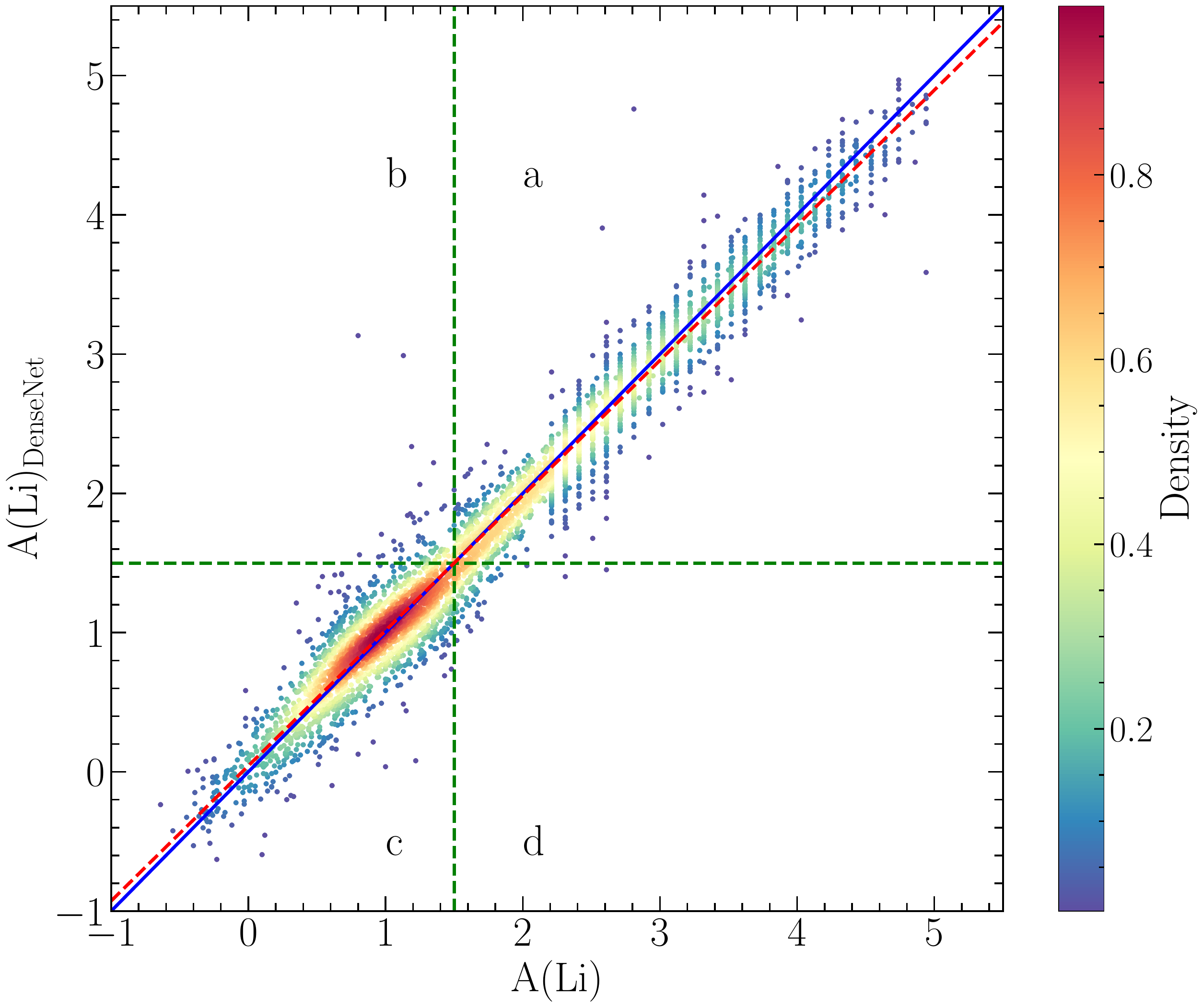}
\caption{Distribution of target values A(Li) vs.\,A(Li) predicted by Coord-DenseNet. Region a and region c indicate normal stars that were identified as normal. region b indicates that normal stars were incorrectly identified as Li-rich giants and region d indicates that Li-rich giants were incorrectly identified as normal stars. The solid line indicates the diagonal line and the red dashed line is the best linear fit to the scattered points.
\label{fig:7}}
\end{figure}

\begin{figure*}[htb]
\centering
\includegraphics[width=0.9\textwidth]{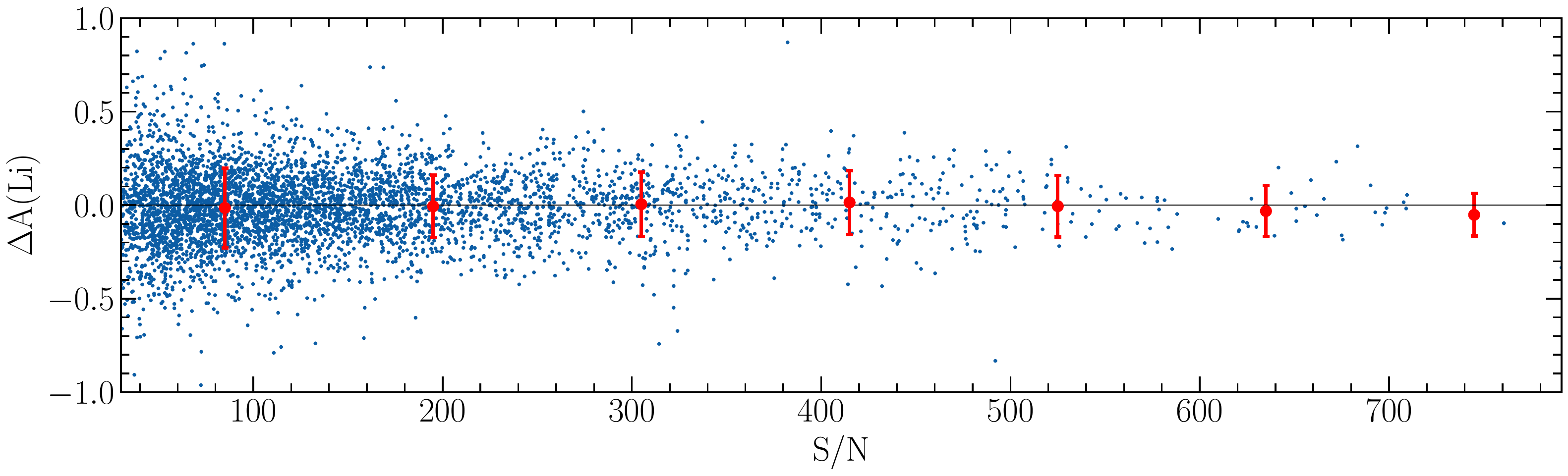}
\caption{Difference in A(Li) between the model predicted values and target values vs. S/N for the test set. The red dots and vertical error bars are the mean and standard deviation, respectively. The size of bins is 100. 
\label{fig:8}}
\end{figure*}

\begin{figure*}[!htb]
\centering
\includegraphics[width=0.92\textwidth]{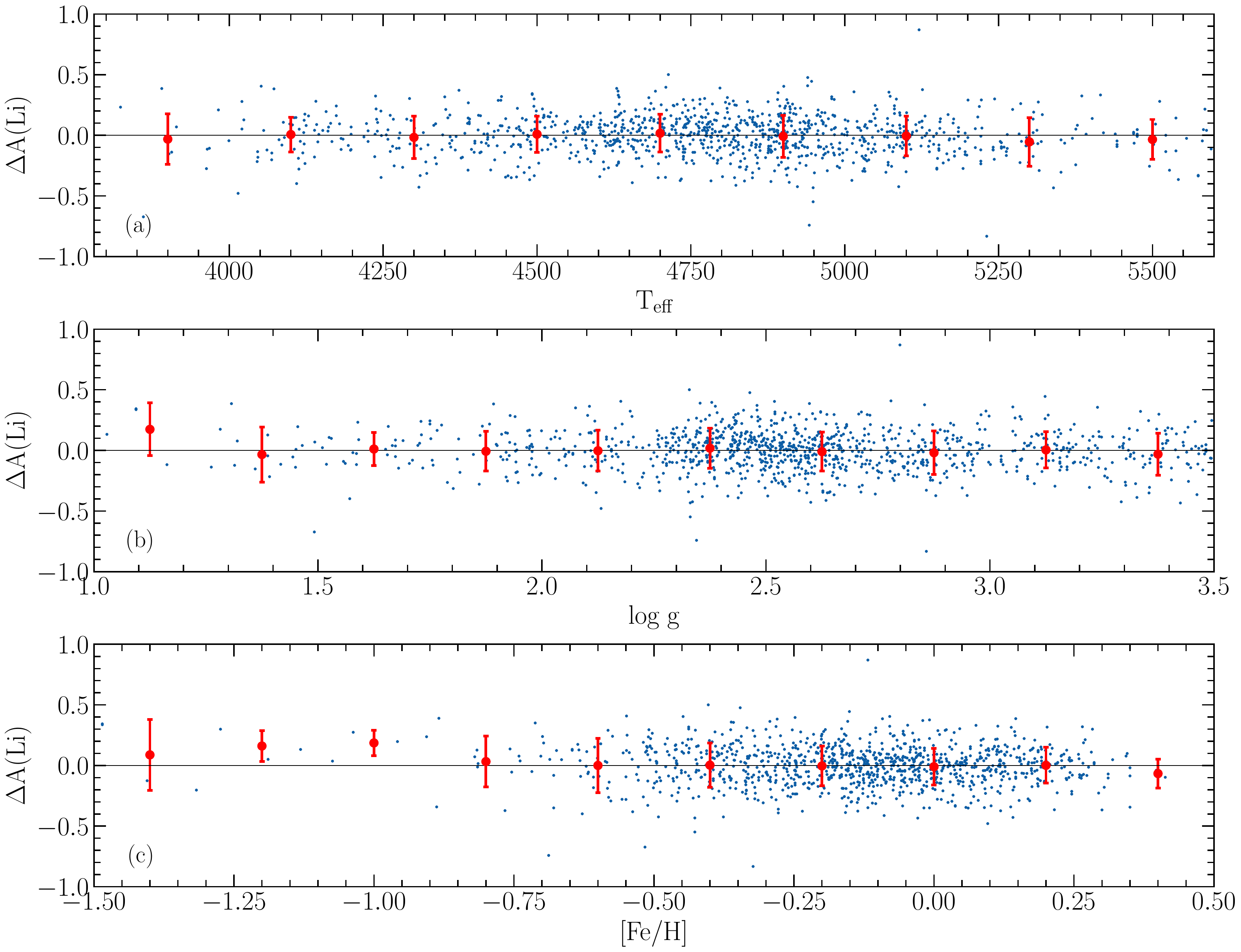}
\caption{The distribution of the difference between model predict values and target values on the test set vs. $\rm T_{eff}\ (a)$, $\rm log\,{g}$\ (b), [Fe/H]\ (c). The bin sizes are 200\,K, 0.25\,dex, and 0.2\,dex respectively. The red dots are the mean value and the error bars are the standard deviation of the differences in every bin.
\label{fig:9}}
\end{figure*}

\section{Experiment}\label{sec:Experiment}
\subsection{Evaluation Method}

The purpose of this paper is to apply known labeled samples to train a deep learning model for Li abundance prediction on LAMOST dr8 low-resolution spectra. Therefore, it is necessary to evaluate the performance of the model. The evaluation indicators of the model performance are as follows:

\begin{itemize}

\item [1)]
Mean absolute error(MAE):

\begin{equation}
MAE=\frac{1}{N} \sum_{i=1}^{N}\left|y_{i}-\hat{y}_{i}\right|
\end{equation}
$N$ is the total number of samples, $y_{i}$ is the label value of the ith sample, and $\hat{y}_{i}$ is the predicted value of the ith sample. It represents the mean of the absolute error between the predicted and observed values
\item [2)]
Standard deviation ($\sigma$):

\begin{equation}
\sigma=\sqrt{\frac{1}{N} \sum_{i=1}^{N}\left(E_{i}-\bar{E}_{i}\right)^{2}}
\end{equation}

where $E_{i}=y_{i}-\hat{y}_{i}$ and $\bar{E}_{i}$ is the mean value of all $E_{i}$. It measures the dispersion of the difference between the predicted and target values of DenseNet.
\end{itemize}

\begin{figure}[hbtp]
\centering
\includegraphics[width=0.46\textwidth]{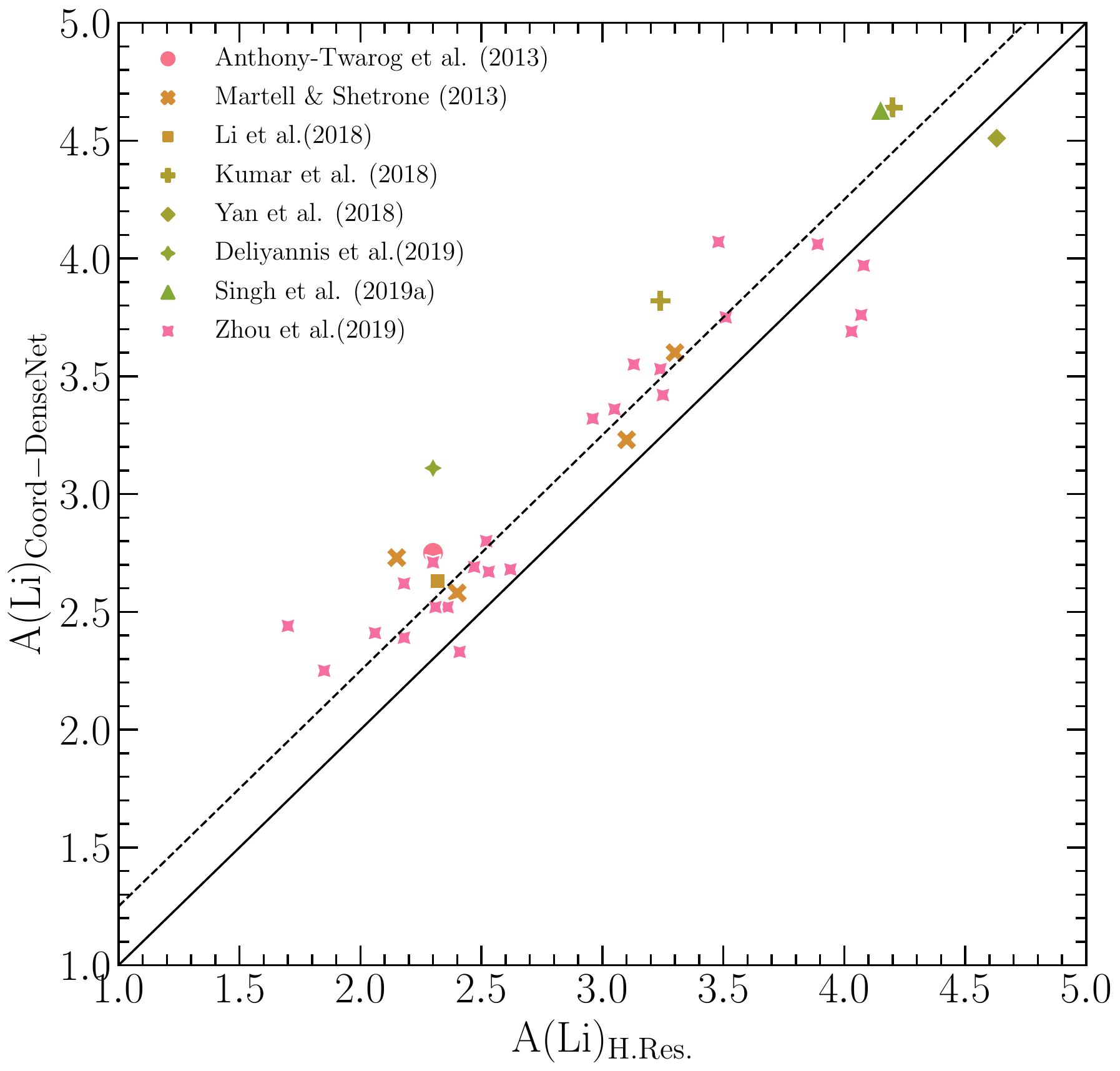}
\caption{ $\rm A(Li)_{Coord-DenseNet}$ of LAMOST LRS DR8 predicted by Coord-DenseNet versus $\rm A(Li)_{H.Res.}$ of the high-resolution literature. The solid line represents the diagonal line, while the black dotted line is the overall shift of 0.27\,dex.
\label{fig:10}}
\end{figure}

\begin{figure}[!htb]
\centering
\includegraphics[width=0.46\textwidth]{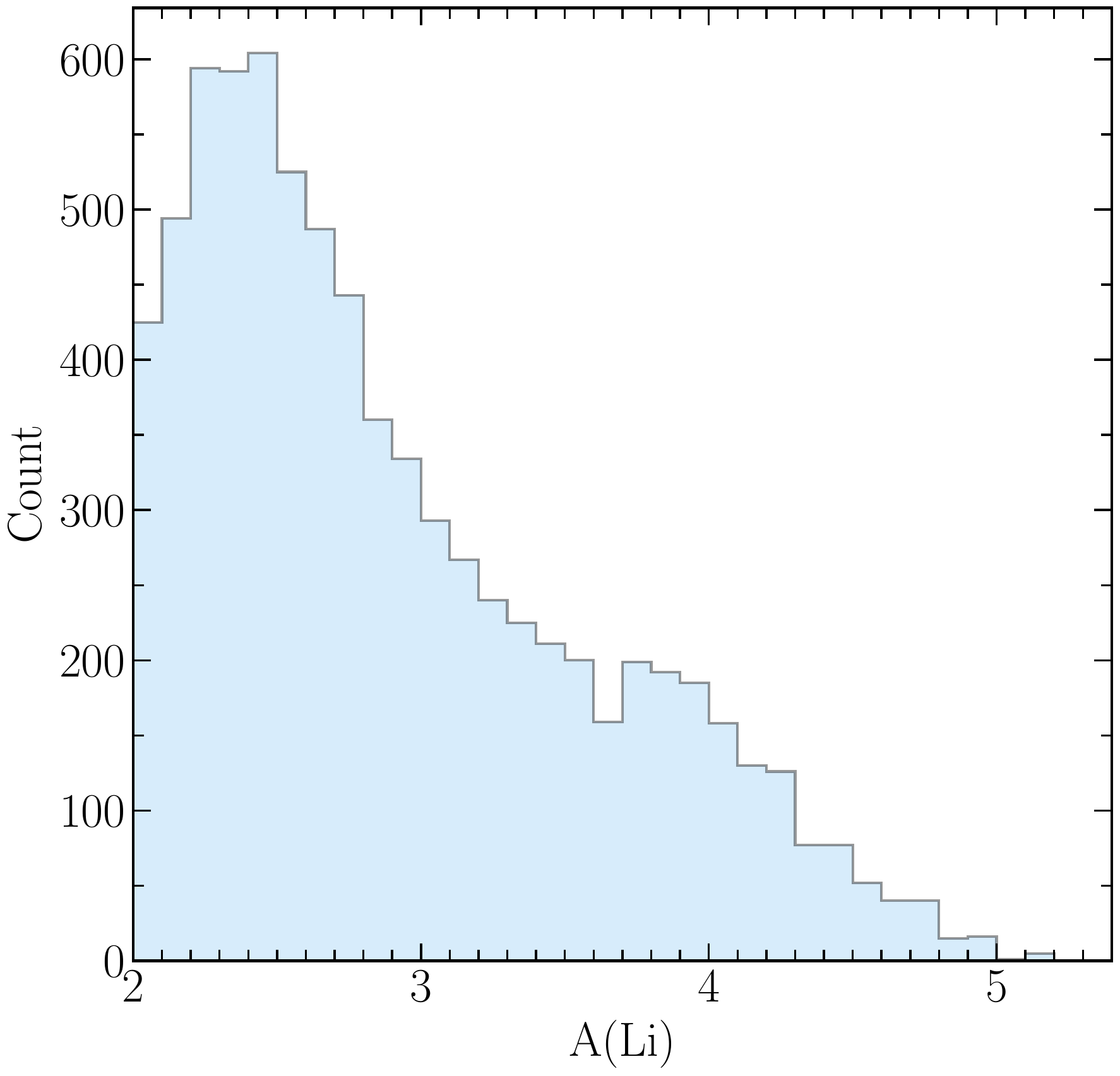}
\caption{Distribution of A(Li) for 7768 Li-rich giants predicted by the model. The bin size is 0.1\,dex. As A(Li) increases, the number of Li-rich giants in each bin decreases, except for the first few bins.
\label{fig:11}}
\end{figure}

\begin{figure*}[htbp]
\centering
\includegraphics[width=0.9\textwidth]{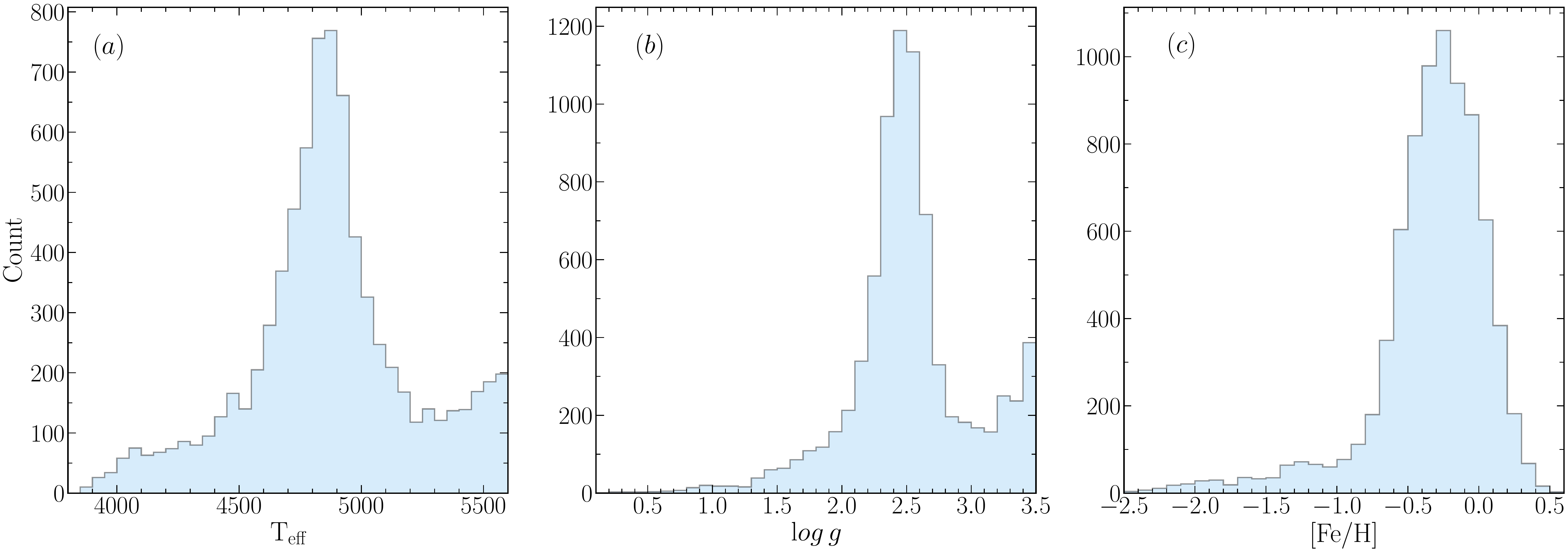}
\caption{The A(Li) versus $\rm{T_{eff}}$ (a), $\rm log\,{g}$ (b), and [Fe/H] (c). The bin sizes are 50\,K, 0.1\,dex, and 0.1\,dex, respectively. In panel (a), there is a clear peak at 4850\,K. When $\rm{T_{eff}}$ $\textgreater$ 5300\,K, the number of Li-rich giants shows an increasing trend with increasing temperature. In panel (b), there is a peak at $\rm log\,{g}$ of 2.5\,dex, and 3.5\,dex, respectively. In panel (c), [Fe/H] has a clear peak near -0.3\,dex.
\label{fig:12}}
\end{figure*}

\begin{figure}[!htb]
\centering
\includegraphics[width=0.46\textwidth]{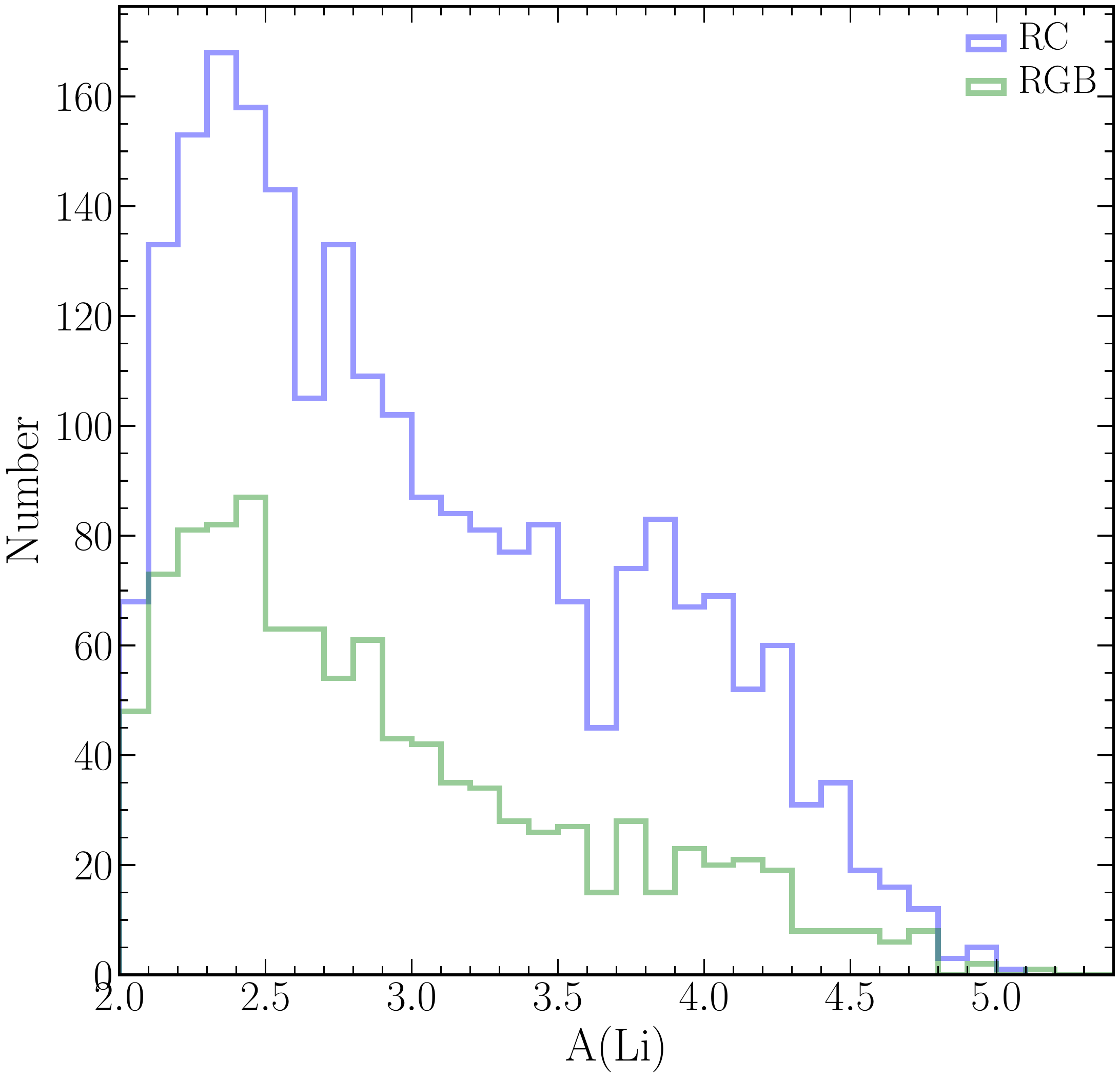}
\caption{The Li abundance distribution of 2323 RC stars (blue) and 1029 RGB stars (green) obtained by crossing our predicted Li-rich giants with Table\,1 of \cite{ting2018large}. The bin size is 0.1 dex. The distribution trends of RC and RGB stars in different A(Li) ranges are consistent.
\label{fig:13}}
\end{figure}



\begin{table*}[!htbp]
\centering
\caption{Information of the Li-rich Giants of Our Sample. \label{table2}}

\begin{tabular}{ccccccccc}
\toprule

\multirow{2}{*}{$\rm LAMOST\ ID$} & \multirow{2}{*}{obs\_id} & R.A. & Decl. & $\rm T_{eff}$ & $\rm log\,{g}$ & [Fe/H] &\multirow{2}{*}{S/N} & A(Li) \\
& &(deg) &(deg) & (K)& (dex) & (dex) & & (dex) \\
\midrule
J000001.30+494500.7 & 250901072 & 0.005424 & 49.7502  & 4443.02 & 2.508 & 0.497  & 129.68 & 2.552614 \\
J000005.50+454110.6 & 370710074 & 0.022935 & 45.68629 & 4803.55 & 2.384 & -0.056 & 307.76 & 3.775815 \\
J000007.78+410505.4 & 281202067 & 0.032427 & 41.08485 & 5259.31 & 3.321 & 0.151  & 143.18 & 2.858763 \\
J000015.86+500614.1 & 182612060 & 0.066101 & 50.10393 & 4729.38 & 2.392 & -0.019 & 58.2   & 2.88857  \\
J000022.92+544825.2 & 269501042 & 0.09554  & 54.80702 & 4908.63 & 2.425 & -0.394 & 97.33  & 4.499752 \\
J000036.02+273038.9 & 492111245 & 0.150102 & 27.51081 & 4959.01 & 2.442 & -0.752 & 184.01 & 2.89567  \\
J000041.35+585002.3 & 269511009 & 0.172322 & 58.83398 & 4690.18 & 2.675 & 0.06   & 185.66 & 3.817895 \\
J000108.96+072932.9 & 496104095 & 0.287347 & 7.492476 & 4734.71 & 2.493 & -0.165 & 208.64 & 3.763314 \\
J000119.92+082335.9 & 496104203 & 0.33209  & 8.393309 & 4804.69 & 2.318 & -0.474 & 375.7  & 2.587063 \\
J000151.65+265848.4 & 492109107 & 0.465241 & 26.98013 & 5070.45 & 2.481 & -0.517 & 268.73 & 4.88001  \\
...&...&...&...&...&...&...&...&...\\ \bottomrule

\end{tabular}
\end{table*}

\subsection{performance on test data}

After model training, to evaluate the performance of the model, we tested it on a test set containing 4496 spectra, and Figure\,\ref{fig:7} shows the distribution of the predicted and target values of the model for A(Li). The results of the test set $\rm MAE= 0.15\,$dex and $\sigma =0.21\,$dex. Li-rich stars with labels A(Li) $\textgreater$ 1.5\,dex are shown in regions a and d, where region a is a correctly identified Li-rich star and region d is a Li-rich giant incorrectly identified by the model as a normal star. Regions b and c are normal stars with labels A(Li) $\textless$ 1.5\,dex, where region b is the region where the model incorrectly identifies normal stars as Li-rich stars, and region c is the region where the model correctly identifies normal stars. As shown in Table\,\ref{table1}, Coord-DenseNet correctly identifies 2170 Li-rich stars with $95.76\,\%$ accuracy, and incorrectly identifies 96 normal stars as Li-rich stars.

Figure\,\ref{fig:8} shows the difference in A(Li) between Coord-DenseNet predicted values and target values of a target as a function of S/N for all 4496 spectra in the test set. From the figure, we can see that the standard deviation of $\triangle \mathrm{A}(\mathrm{Li})$ decreases from 0.2\,dex to 0.1\,dex as S/N increases, which indicates that the random error is very sensitive to S/N and the higher the S/N the smaller the error, which is in accordance with our expectation.
Panel (a) to panel (c) in Figure\,\ref{fig:9} shows the relationship between $\triangle \mathrm{A}(\mathrm{Li})$, and the three atmospheric parameters\,($\rm T_{eff}$, $\rm log\,{g}$, and [Fe/H]), respectively. To avoid the influence of S/N, only the 1107 spectra with S/N $\geq$ 200 are included in the figure. In panels (a) and (b), the standard deviation does not show a significant correlation with both $\rm T_{eff}$ and $\rm log\,{g}$, while in panel (c) the sample size of the fraction [Fe/H] $\textless$ -0.7 is sparse and insufficient to analyze the correlation between $\triangle \mathrm{A}(\mathrm{Li})$ and [Fe/H]. For the fraction [Fe/H] $\textgreater$ -0.7, the standard deviation slowly decreases as [Fe/H] increases.


\section{Results and Discussion\label{sec:Discussion}}

\subsection{Predictions for LAMOST LRS DR8 Spectra}

The LAMSOT LRS DR8 includes all low-resolution spectral data observed between October 24, 2011, and May 27, 2020. The conditions
{$\rm T_{eff}\leq5600\,K$}, $\rm log\,{g}\leq3.5$, $\rm S/N\geq30$, and no missing redshift were used, which are the same constraints as the data used for model training. We then obtained 942148 spectra from LAMOST LRS DR8 corresponding to 758208 giants. After the same preprocessing process as the training data, we made predictions of the A(Li) of these spectra using the Coord-DenseNet.

\subsection{Comparison with the Previous Literature}

To validate our method used for calculating A(Li), we compared A(Li) derived by us with those which have been analyzed by other pieces of literature.

\citet{mallik2003lithium} determined the Li abundance in 127 F and G Pop\,\uppercase\expandafter{\romannumeral1} stars based on measurements of the equivalent width of the\,$\lambda 6707$\,\AA \,$ \mathrm{Li\,\uppercase\expandafter{\romannumeral1}}$ line from high-resolution CCD spectra. \citet{takeda2005lithium} determined the Li abundances of 160 neighboring F-K dwarfs/subgiants in the Galactic disk by profile fitting analysis at the Li(+Fe) 6707-6708\AA\, feature. \citet{prisinzano2007lithium} calculated the Li abundance of open cluster NGC3960 from VLT/FLAMES observation by the lithium equivalent width. \citet{da2009search} calculated the local thermodynamic equilibrium\,(LTE) Li abundance for 376 stars using measurements of the equivalent width of the lithium resonance line. \citet{takeda2010behavior} investigated the correlation between Li abundance and stellar rotation in 118 solar analogs.  \citet{gonzalez2010parent} determine Li abundances and $\upsilon\, {\rm sin} i$ values from new spectra of 53 stars with Doppler-detected planets. \citet{mishenina2012activity} used synthetic spectroscopy to calculate the Li abundance of 150 slowly rotating stars located in the lower part of the main sequence belt and analyzed the relationship between stellar parameters and Li abundance. \citet{anthony2013lithium} find a Li-rich red giant below the clump in the Kepler cluster, NGC 6819. \citet{martell2013lithium} has identified 23 post-turnoff stars with A(Li) greater than 1.95\,dex.  \citet{delgado2014li} determined the Li abundance of 326 main sequence stars. \citet{liu2014lithium} derived the Non-local thermodynamic equilibrium\,(N-LTE) Li abundances for a total of 378 G/K giant stars from the Okayama Planet Search Program and the Xinglong Planet Search Program. \citet{zhao2016systematic} did a study to determine the N-LTE abundances of 17 chemical elements from Li to Eu in field stars near the Sun.   \citet{li2018enormous} identified 12 metal-poor stars with lithium enrichment from the low-mass stars of the Milky Way Halo. \citet{anthony2018li} determined the Li abundance of 85 G-K stars in Open Cluster (OC) M35. \citet{kumar2018two} reported two new super Li-rich K giants, KIC2305930 and KIC12645107. \citet{yan2018nature} reported a Li-rich giant with an A(Li) of 4.51\,dex. \citet{deliyannis2019vizier} use spectroscopy of 333 NGC 6819 stars and Gaia astrometry to map Li evolution from the giant branch tip to 0.5 mag below the Li dip. \citet{zhou2019high} used high-resolution observations to identify 44 new Li-rich giants from the LAMOST survey. \citet{singh2019spectroscopic} reported the discovery of two new super Li-rich K giants: HD 24960 and TYC 1751-1713-1. \citet{romano2021gaia} used data from the Gaia-ESO survey to obtain Li abundances for 26 OCs and star-forming regions with ages ranging from young ($\sim$3 Myr) to old ($\sim$4.5 Gyr). \citet{magrini2021gaia} obtained the Li abundances of 4212 cluster stars and 7369 field stars from Gaia iDR6.

We collected the A(Li) published in the above literatures for comparison with the  A(Li) predicted by our model. Finally, we obtained a literature reference set containing 42,659 stars. However, most of them were not observed by LAMOST and we crossed to only a small fraction of them. The results of the $\rm A(Li)_{Coord-DenseNet}$ predicted by Coord-DenseNet and the $\rm A(Li)_{H.Res.}$ given in the literature reference set crossed to a total of 35 stars are shown in Figure\,\ref{fig:10}. The scatter is 0.25\,dex and the MAE is 0.32\,dex. We can see that there is an overall offset of 0.27\,dex between $\rm A(Li)_{Coord-DenseNet}$ and $\rm A(Li)_{H.Res.}$, the reasons for the deviation may be as follows: First, the labels for our training data come from \citetalias{gao2019lithium}, and the A(Li) predicted in \citetalias{gao2019lithium} is about 0.09\,dex higher than those in the high-resolution literature. Second, when we compared the consistency of A(Li) labels of common stars from \citetalias{gao2021lithium} and \citetalias{gao2019lithium} in Figure\,\ref{fig:fig1} of Section \ref{sec:DATA}, $\rm A(Li)_{M.Res.}$ is overall 0.18\,dex larger than the $\rm A(Li)_{L.Res.}$. We use Eq.\ref{eq:1} to increase the $\rm A(Li)_{L.Res.}$ label of the positive samples by about 0.18\,dex.  It is not ruled out that the $\rm A(Li)_{M.Res.}$  is on the high side. We notice that in Figure\,\ref{fig:6}\,(a) of \citetalias{gao2021lithium}, they compared the Li abundances calculated by their method\,($\rm A(Li)_{LAMOST}$) with those determined from GALAH\,($\rm A(Li)_{GALAH}$) and the consistency was good. However, they recalculated the $\rm A(Li)_{M.Res.}$ by using the atmospheric parameter\,($\rm T_{eff}$, $\rm log\,{g}$, [Fe/H]) from GALAH. When we compared their original $\rm A(Li)_{M.Res.}$ with those from GALAH, the result showed that $\rm A(Li)_{M.Res.}$ was generally higher.

\subsection{LAMOST LRS DR8 Catalog of Stellar Parameters and Li Abundance}

Based on the definition of Li-rich giants\citep{liu2014lithium,casey2016gaia,magrini2021gaia}: A(Li)\,>=\,2.0\,dex, \,$\rm{T_{eff}}$\,<=\,5600\,K, $\rm log\,{g}$\,<=\,3.5\,dex, we recognized 9126 spectra corresponding to 7768 Li-rich giants from the LAMOST DR8 LRS, accounting for about 1.02\% of all giants. This ratio is in good agreement with $\sim 0.5\,\%-2\,\%$ of all Li-rich giants reported in other literature\citep{brown1989search,charbonnel2000nature,kumar2011origin,ruchti2011metal,martell2013lithium,smiljanic2018gaia,reddy2019study}. For more than one A(Li) of the same giant star, we use the predicted value of the spectra with the highest S/N. The information of these 7768 Li-rich giants is listed in Table\,\ref{table2}, including the identifier for
the corresponding star(LAMOST ID), the LAMOST spectrum identifier(obs\_id), coordinate information(R.A., Decl.), stellar atmospheric parameters($\rm T_{eff}$, $\rm log\,{g}$, [Fe/H]) provided by LAMOST Stellar Parameter Pipeline (LASP; \citealt{luo2015first}), r-band signal-to-noise ratio(S/N) and A(Li) predicted by Coord-DenseNet. It should be noted that due to the lack of sufficient training data, the portion of our predictions with A(Li)\,\textgreater\,4.5\,dex may have extrapolation and their abundance values are for reference only.

Figure\,\ref{fig:11} shows the histogram of the number of A(Li) with abundance for the 7768 Li-rich giants. It can be seen from the figure that overall the number of corresponding Li-rich giants decreases continuously as the A(Li) increases. It can not be ignored that the number of Li-rich giants with A(Li) in the range of 2.0\,$-$\,2.2\,dex is significantly less. Checking the model training label sample in Figure\,\ref{fig:fig1}, we found that the number of samples in this part was very small. The reason may be that although we performed data augmentation for this part, the model still failed to fully learn the features possessed by the spectra in this abundance range because of the data imbalance, so the prediction for this part was not as good as it should be. Another possible reason for this is that as A(Li) decreases, the $\mathrm{Li\,\uppercase\expandafter{\romannumeral1}}$ line also becomes weak, which also makes it difficult for the model to give accurate predictions.

Of these 7768 Li-rich giants, 4049 have Li abundances above the primordial value of A(Li) = 2.7\,dex, and these stars are called super Li-rich giants, which account for about 52\% of all Li-rich giants. This percentage is 25.6\% in \citet{martell2021galah}, but it should be noted that the definition of Li-rich in their paper is A(Li)\,\textgreater\,1.5\,dex. Considering that our predicted A(Li) has a deviation of 0.27\,dex compared to $\rm A(Li)_{H.Res.}$, Martell's A(Li)\,=\,1.5\,dex corresponds to our A(Li)\,=\,1.77\,dex, and their A(Li)\,=\,2.7\,dex corresponds to our A(Li)\,=\,2.97\,dex. Taking this deviation into account, the ratio of super Li-rich giants in Li-rich giants is about 23\%, which is very close to the percentage given by \citet{martell2021galah}. 

Figure\,\ref{fig:12} shows the histograms of the number of our Li-rich giant sample versus $\rm T_{eff}$, $\rm log\,{g}$, and [Fe/H], respectively. For the histogram of temperatures, there is a maximum peak near 4850\,K. When $\rm {T_{eff}}$ $\textgreater$ 5300 K, the number of Li-rich giants keeps increasing with the increasing temperature. there are two peaks near $\rm log\,{g}$ 2.5\,dex and 3.5\,dex, the vicinity of the first peak corresponds to many RC stars. The distribution of metallicity shows that there is a peak at [Fe/H] about $-\,0.3$\,dex.

Recent studies have found that RC stars represent a higher proportion of lithium-rich giants than RGB stars: the percentage is 58\% in \citet{martell2021galah} and 86\% in \citet{yan2021most}. \citet{ting2018large} separated the RC stars from the RGB stars based on LAMOST DR3 stellar spectra and released a catalog containing 149,732 RC stars and 197,995 RGB stars in their Table 1. Figure\,\ref{fig:13} shows that we crossed our identified Li-rich giants with this catalog and obtained 2323 RC stars and 1029 RGB stars, with RC stars accounting for about 69.3\% of the Li-rich giants, which is consistent with the literature.



\section{Summary}\label{sec:Summary}

In this paper, we develop a data-driven model based on convolutional neural networks named Coord-DenseNet to evaluate the A(Li) from LAMOST DR8 LRS, the error of the Coord-DenseNet model on the test set is about 0.2\,dex. Using the model, we identified 7768 Li-rich giants with A(Li)$ \geq 2$\,dex from LAMOST LRS DR8, accounting for about $1.02\,\%$ of all giants. These Li-rich giants further enrich the sample pool of known Li-rich giants and will help to better study the variation pattern of Li during stellar evolution. 

This work is supported by the National Natural Science Foundation of China (NSFC) under grant Nos.11803016, U1931209 and 11873037.

\bibliographystyle{aasjournal}
\bibliography{bibtex}

\end{document}